\documentstyle[multicol,aps,prb,epsf]{revtex}
\draft
\begin{document}
\title{\hspace{-1 cm}{(\small \it to appear in Phys. Rev. Lett.)}\\
 \large 
Photoabsorption  Spectra of Na$_n^+$ clusters: Thermal Line-Broadening Mechanisms
}
\author{ M. Moseler$^{1,2}$, 
                H. H\"akkinen$^2$, 
            and Uzi Landman$^2$,\\
$^1$Theoretische Quantendynamik, Fakult\"at f\"ur Physik, 
Universit\"at Freiburg,  
79106 Freiburg, Germany\\
$^2$School of Physics, Georgia Institute of Technology,
Atlanta, GA 30332-0430}
\date{\today}
\maketitle
%
%ABSTRACT
%
\begin{abstract}
Photoabsorption cross sections of small sodium cluster cations
(Na$_n^+$, $n$$=$3,5,7 and 9)
were calculated at various  temperatures with the time-dependent 
local-density-approximation (TDLDA)   in conjunction with 
{\it ab initio} molecular dynamics simulations, yielding 
spectra that agree with measured ones without ad-hoc line broadening 
or renormalization. Three thermal 
line-broadening mechanisms are revealed: (I) lifting of level degeneracies
 caused by symmetry-breaking ionic motions, (II) oscillatory  
shifts of the entire spectrum caused by breathing vibrations, and
 (III) cluster structural isomerizations. 
\end{abstract} 
\begin{multicols}{2}
\narrowtext
%
%
%Introduction
%
%
Optical spectroscopy provides invaluable insights
into the electronic structure, ionic configurations, 
thermal processes and dynamics in metal clusters, as 
well as about the size-dependent evolution of these 
properties. Indeed, investigations of these issues 
have been pursued rather intensively for over a decade
both experimentally with photodepletion spectroscopy~\cite{expaut,Schmidt} and 
theoretically~\cite{Ekardt,Pacheco1,Yannouleas,Bertsch,Bonacic,Rubio,Pacheco2}
with the main
methodology  employing the 
TDLDA~\cite{Ekardt,Yannouleas,Casida}
in conjunction with either jellium models~\cite{Ekardt,Pacheco1,Yannouleas}
(including allowance for shape-deformations)
 where the ionic 
background is smeared out uniformly, or with electronic 
structure calculations of various degree of sophistication
where the discrete nature of the ions is incorporated
accurately~\cite{Bonacic,Rubio} or 
perturbatively~\cite{Pacheco2}. 
While valuable information pertaining to the optical
excitations and damping mechanisms in metal clusters
has been obtained through such studies, including
the sensitivity of  optical features to  cluster 
geometries, a first-principles theoretical description of 
the optical line-shapes, the absolute absorption cross-sections,
 the relevant microscopic line-broadening mechanisms and their 
thermal variations, is lacking. This is reflected in the
 common ad-hoc employment of line broadening through convolution of 
spectral lines calculated for selected static cluster 
configurations with Gaussian or Lorentzian
 functions~\cite{Ekardt,Pacheco1,Yannouleas,Bonacic,Rubio},
or (multiplicative) renormalization of the calculated spectra to the measured 
ones~\cite{Pacheco2,Kresin}. 

In this letter, we demonstrate that photoabsorption cross sections,
$\sigma(\omega)$, calculated via the TDLDA along finite-temperature
Born-Oppenheimer (BO) local-spin-density (LSD) molecular dynamics (MD)
(BO-LSD-MD~\cite{Barnett}) phase-space trajectories, provide a
{\it quantitative ab-initio description} of the {\it absolute 
magnitudes, peak positions, and line shapes} of optical absorption spectra 
measured at various temperatures from Na$_n^+$ ($n$$=$3, 5, 7 and 
9) clusters~\cite{Schmidt}. Furthermore, analyses of the results,
and in particular  the temporal correlations between the calculated
$\sigma^t(\omega)$ (where the superscript denotes the value at instant
 $t$) and the dynamically evolving cluster configurations, 
reveal three  thermally induced mechanisms which govern 
the spectral characteristics: (I) symmetry breaking of the
optimal (0 K) cluster geometries
due to thermal motion of the ions, leading to lifting of electronic
level degeneracies and resulting in splitting (fragmentation) of
spectral lines; (II) symmetry conserving breathing motions 
that modulate the effective density, resulting in oscillatory
 frequency shifts of the entire spectrum 
("spectral sweeping") that influence the optical absorption peak 
positions and line width; (III) opening of the accessible
 configurational space, resulting in structural isomerization
and consequent electronic structure and spectral variations.
The first two mechanisms are operative for  smaller 
clusters (Na$_3^+$ and Na$_5^+$) even at relatively low 
temperatures (e.g. 100 K), while all three occur for 
the larger ones.
Furthermore, the thermal dependence of the spectral characteristics
is analyzed, resulting in remarkable first-principles agreement between the 
calculated and measured optical absorption cross sections.

%
%
%Methods
%
%
The Kohn-Sham (KS) equations with
generalized gradient corrections (GGA)~\cite{Perdew}
and non-local  pseudopotentials~\cite{Troullier} were
solved for Na$_n^+$ clusters ($n$$=$3, 5, 7, and 9), and 
based on  the Hellman-Feynman forces, finite temperature 
 trajectories were generated by Langevin dynamics through the 
use of the BO-LSD-MD (with GGA) method~\cite{Barnett}.
For each cluster size, starting 
from the  (presumed~\cite{Bonacic}) GGA-optimized ground state (GS) 
geometry and 
an equilibration of 
10 ps duration, the system was allowed to evolve dynamically at 
the prescribed  temperature for an additional period 
 of $t_s$$=$10 ps.
At each instant $t$, 
the excitation 
energies $\omega_I^t$ were determined by solving the TDLDA eigenvalue
problem~\cite{Casida} 
\[\sum_{kl}( (\epsilon^t_{ij})^2 \delta_{ik}\delta_{jl}+
  4\sqrt{\epsilon^t_{ij}\epsilon^t_{kl}}K^t_{ij,kl}
)F^t_{I,kl}=(\omega^t_I)^2F^t_{I,ij},\]
where  $i,j$ and $k,l$ denote particle-hole pairs~\cite{Conv}, 
and  $\epsilon^t_{ij}$ the corresponding KS orbital  energy differences.
The coupling matrix is given by
\[
K^t_{ij,kl}\hspace{-0.1 cm} =
\hspace{-0.15 cm}\int \hspace{-0.15 cm} d^3r_1 d^3r_2\, \phi^t_{ij}({\bf r}_1) 
[\frac{1}{r_{12}}+\delta ({\bf r_{12}})
\frac{\partial v_{xc}(\rho^t({\bf r}_1))}{\partial \rho^t({\bf r}_2)}]
\phi^t_{kl}({\bf r}_2),\]
with the GS electron density $\rho^t$,
the product of two (real) KS orbitals $\phi^t_{ij}$, 
and the LDA exchange-correlation potential $v_{xc}$ .
The  oscillator strength (OS) 
\begin{equation}
f^t_I=\frac{4}{3}\sum_{\nu=1}^3\bigg|\sum_{ij} 
\int d^3r\phi^t_{ij}({\bf r}) r_\nu 
\sqrt{\epsilon^t_{ij}}F^t_{I,ij}\bigg|^2\label{f_I}\label{oszistren}
\end{equation}
corresponding to the transition energies $\omega^t_I$,
 were accumulated in 
bins of width $\Delta \omega$$=$$0.025$ eV
yielding the
absolute photoabsorption cross section per valence electron
\[
\sigma(\omega)=\frac{1}{n-1}
\frac{2\pi^2}{c}\frac{1}{t_s}\int_0^{t_s}dt\frac{1}{\Delta \omega}
 \sum_{I, \omega^t_I \epsilon [\omega,\omega+\Delta \omega]} f^t_I.
\]

%
% 
%
%
%Results
%
%
We note first the overall 
good agreement between the magnitudes
and shapes of
the calculated and measured photoabsorption
 cross sections (shown here for Na$_n^+$, $n$$=$3,5 and 9); 
compare the red histograms with the black curves in 
Fig.~1, and note that  neither  (artificial) numerical broadening
 nor 
a renormalization of the spectra has been applied.
The slight deviations in  peak positions 
are within the known accuracy of the TDLDA ($\pm$0.2 eV)~\cite{Casida}.

The calculated Na$_3^+$ spectrum 
 at $T$$\approx$100~K (Fig.~1a) exhibits a low energy 
peak originating~\cite{OrbContrib} from two transitions 
$\omega^t_1$=$\omega^t_2$$\approx$2.65 eV,
 from the occupied s-like orbital
 to two empty p-like  
KS orbitals 
located in the plane of  the Na$_3^+$ equilateral triangle
(see grey insets 1, 2 and  3 at the left of Fig.~1a).
The high energy peak at $\omega^t_3$$=$3.41 eV 
is due to an excitation to another 
p-like orbital that is   perpendicular to the cluster 
plane (grey inset 4 in Fig.~1a). 
In the experiments $\sigma(\omega)$ is determined from  
the depletion of the Na$_n^+$ intensity  due to dissociation
 following absorption of a photon $\hbar \omega$. 
Excitation into the {\it in-plane} anti-bonding orbitals (insets 2 and 3)
promotes such dissociation and is accompanied by  exhaustion of the 
OS for the  first (lower energy) measured peak.
Excitation into the {\it out-of-plane} orbital (inset 4)
has no direct destabilizing effect and consequently
in the measurements  only 66\% of the Thomas-Reiche-Kuhn sum rule (TRK)~\cite{Casida}
is observed while the full 
  TRK sum rule is found in the theoretical spectrum
shown in Fig. 1a (as well as  in  all the calculated 
spectra that we show here).   

Thermal motions distort the   D$_{3h}$ symmetry of the Na$_3^+$ GS 
and  the degeneracy of the low-energy transitions $\omega_1$ and
$\omega_2$  is lifted. Indeed, the  evolution
of  $\delta \omega_{21}^t$$=$$\omega^t_2$$-$$\omega^t_1$ 
coincides with the  temporal behavior
of the standard deviation $\Delta^t_b$$=$$\sqrt{\sum_i 
(b^t_i-{\bar b} ^t)^2}/3$
of the three bond distances $b^t_i$
 (see Fig.~2a). 
Additionally,  the mean bond distance ${\bar b} ^t$ 
  is strongly anti-correlated 
with the mean frequency 
$\bar \omega_{12}^t$$=$$(\omega^t_1+\omega^t_2)/2$  as
well as with the position of $\omega^t_3$ 
(compare the solid line with the lower and upper dashed curves in Fig.~2b) -
that  is,
 an increase of the ionic density (decrease of ${\bar b} ^t$)
results in a  
blue shift of the entire spectrum.  Thus, the spectral line width of  Na$_3^+$
 can be fully understood
in terms of
line-splitting and spectral sweeping 
with the first caused by 
symmetry breaking (degeneracy 
%
%     FIG1
%
\begin{figure}[tb]
\setlength{\unitlength}{1cm}
\epsfxsize=7.5cm
\epsfbox{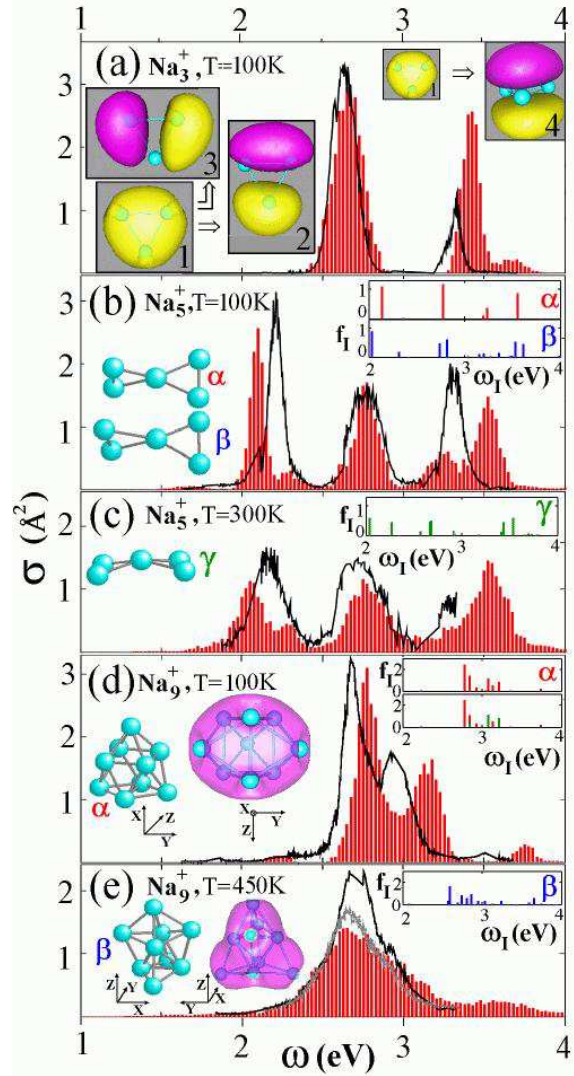}
\caption[]{Theoretical (red histograms) and experimental~\cite{Schmidt} (black curves)
photo absorption cross section for  Na$_3^+$ at 100 K (a),  
Na$_5^+$ at 100 K (b) and 300 K (c), as well as for 
Na$_9^+$ at 100 K (d) and at 450 K (e). Grey insets
in (a) display isosurfaces of the 4 lowest KS orbitals where yellow and purple
distinguish the sign of the wave function  and light blue 
spheres mark the position of the ions. 
The right top corner insets in (b-e) show static TDLDA spectra:
(b) for the Na$_5^+$ GS (structure $\alpha$), and
for an instantaneous  structure $\beta$ 
with an elongated left triangle; (c) for an instantaneous 
bent planar  geometry ($\gamma$).
The calculated spectrum shown in (d) derives 
exclusively from configurations lying in the basin of the 
GS structure $\alpha$ whose 
oblate spheroidal electron density is depicted by 
the purple isosurface.
The static spectrum of $\alpha$ is shown at the top inset in (d), 
and in the bottom inset we  show the decomposition
of that  spectrum into the component along the z-axis (green sticks) 
and the perpendicular components (red sticks).
The  Na$_9^+$ isomer  $\beta$ is shown in (e),
along with its  electron density (purple isosurface)
 and its static spectrum (right top inset).
To highlight the sensitivity  of the experimental results to the 
temperature, we include in (e) the measured  spectrum~\cite{Schmidt} 
at 540 K (grey curve). 
}
\end{figure}
\noindent lifting, mechanism I) and the latter by
breathing vibrations of the cluster (density oscillations, mechanism II).

Similarly, the  finite temperature dynamics of  Na$_5^+$
distorts (even at $T$$=$100 K)  the  optimal D$_{2d}$ GS symmetry
(see GS $\alpha$ in Fig.~1b where the perpendicular left and right 
triangles are equivalent, and compare to 
the instantaneous structure $\beta$ shown in Fig.~1b with an elongated 
left triangle), resulting in
fragmentation (I) of  the GS  line at $\omega$$=$ 2.8 eV into
two separate spectral lines $\omega_u$ and $\omega_d$
~\cite{Na5orb} (compare red and blue lines in the up-right 
inset of Fig. 2b).
Consequently, the  temporal variation of the absolute difference 
of the
 average bond lengths
in the left and the right triangle  
($\Delta^t_{lr}$$=$$|{\bar b}^t_l-{\bar b}^t_r|$, serving as a measure 
of the thermal structural distortion) correlates
directly with the   spacing 
$\delta_{ud}^t$$=$$\omega^t_u$$-$$\omega^t_d$ between the frequencies 
of the spectral fragments  (see solid and dashed curve in Fig.~2c).

While the relatively moderate structural distortions occurring 
at 100 K are not sufficient to lift the degeneracy of the states
associated with the low energy absorption line of Na$_5^+$ (at  $\sim$2 eV,
compare insets $\alpha$ and $\beta$ at the upper right of
 Fig. 1b) which remains relatively 
sharp, increase of the temperature to 300 K results in bent configurations
(see instantaneous structure $\gamma$ on the left in Fig. 1c) where the low energy line is 
fragmented (see inset $\gamma$ at the upper right  of Fig. 1c), resulting 
in enhanced broadening of that spectral line (compare Figs. 1b and 1c).
 Further broadening of 
the lines is caused by the aforementioned breathing modes (spectral sweeping mechanism II)
reflected in the direct correlation 
between the interionic Coulomb repulsion $E^t_{coul}$ (as a
measure of  the ionic density)
 and the average
excitation energy ${\bar\omega} ^t$$=$$\sum_I f^t_I\omega^t_I/(n-1)$ 
shown in Fig~2d for the $T$$=$ 300~K simulation. 

For larger clusters, the occurrence of thermal isomerizations opens
an additional line-broadening channel (III)~\cite{Na7}.
The GS structure of the   Na$_9^+$ cluster~\cite{Na9GS} is 
shown at the left of Fig. 1d (marked $\alpha$) and the TDLDA
spectrum calculated for this static configuration exhibits several 
absorption lines distributed in a bimodal-like 
 manner (see top inset at the upper right 
of Fig. 1d). This bimodality, which is in contradiction to the single-line
spectrum predicted by the jellium model~\cite{Yannouleas},
originates from the "oblate shape" of the cluster with radii 
$R_{z}$$=$2.68 \AA{} along the D$_{3h}$ symmetry axis (z in Fig. 1d)
and $R_{xy}$$=$3.22 \AA{} in the xy plane (these radii were determined from the 
diagonalized moment of inertia of the ionic GS structure $\alpha$).
The total electron density
(see purple isosurface in Fig.~1d) is  almost an ideal oblate spheroid.  
A decomposition of the OS of the D$_{3h}$ GS
 into a z ($\nu$$=$3 term in eq. \ref{oszistren}) 
and the xy contribution 
($\nu$$=$1 and 2 terms)
shows clearly that, as expected~\cite{Ekardt}, 
 the smaller spatial dimension in the z direction  
is associated with a  higher excitation 
energy (compare green and red line in bottom inset at
 the upper right of Fig. 1d).
The bimodal character of the 100 K photoabsorption spectrum 
is evident in both the measured and calculated spectra
(Fig. 1d), with further broadening caused by thermal motions 
through line-fragmentation 
%
%
%
%     FIG2
%
\begin{figure}
\setlength{\unitlength}{1cm}
\epsfxsize=7.5cm
\epsfbox{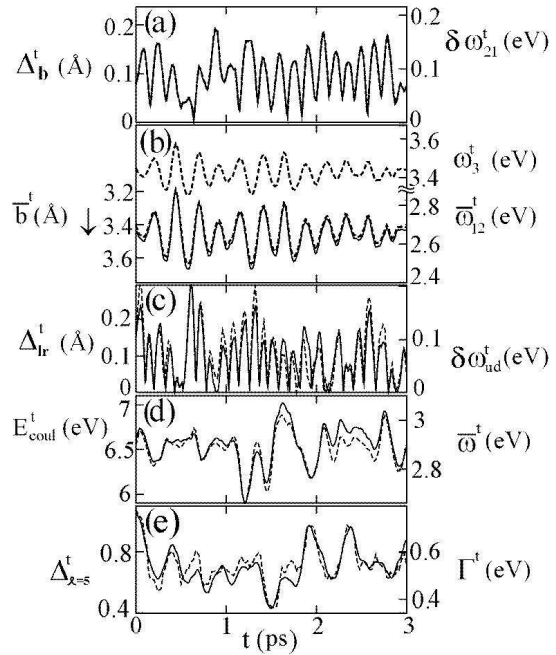}
\caption[]{Evolution of structural and spectral quantities with time:
(a) the  standard 
deviation $\Delta^t_b$ of the three Na$_3^+$ bonds (solid curve) 
coinciding with $\delta \omega^t_{21}$ (dashed curve) at 100 K;
(b) the mean bond length $\bar b^t$ of Na$_3^+$ (solid curve) and the mean 
frequency $\bar \omega_{12}^t$ (lower dashed curve),
as well as the transition energy $\omega_3^t$
(upper dashed curve) at 100 K;
(c) absolute difference $\Delta^t_{lr}$$=$$|\bar b^t_l-\bar b^t_r|$ between the
 average bond length in the left and the right triangle of Na$_5^+$
(solid curve) and $\delta\omega^t_{ud}$ (dashed curve) at 100 K;
(d) Interionic Coulomb repulsion  per ion in Na$_5^+$ (solid line) and mean energy
of the corresponding TDLDA spectrum $\bar \omega^t$ (dashed curve) at 300 K;
(e) $\Delta^t_{\ell =5}$ (solid line) and  $\Gamma^t$ (dashed line) for Na$_9^+$ at 450 K.
}
\end{figure}
\noindent and breathing vibrations of the cluster
(mechanisms I and II).

The main thermal effect  observed in the spectra of Na$_9^+$ 
is the conversion of the low temperature (e.g. $T$$=$100 K, Fig. 1d)
bimodal spectrum to one with a single broad maximum at higher 
temperatures (e.g. 450 K, Fig. 1e). This change in the spectrum 
is caused mainly by transformations between the GS geometry 
($\alpha$, Fig. 1d) and the structural isomer $\beta$ (see configuration 
on the left of Fig. 1e). 
The static spectrum of the $\beta$ isomer is shifted to lower energies with respect 
to that of the GS ($\alpha$) structure (compare top insets marked $\alpha$ and $\beta$
at the upper right corners of Figs. 1d  and 1e). Thus, the high-temperature
broad spectral feature (Fig. 1e) is due to contributions from both  
the $\alpha$ and $\beta$  isomers to the phase-space trajectories generated 
by  the Langevin MD simulation 
at 450 K~\cite{boltzman,sweeping}.

The $\beta$-isomer may be characterized as having an octupolar shape, reflected 
in the  shape of the electron density shown in Fig. 1e (see purple isosurface).
To quantitatively analyze the spectral broadening caused by transitions (at high T)
between the GS ($\alpha$) and the $\beta$-isomer (which involve changes in 
shape-multipolarity) we calculate  the dimensionless multipole
 shape parameters~\cite{Koskinen}, 
$a_{\ell m}$$=$$\frac{\sqrt{4\pi}}{3 r_s^\ell n^{\ell/3+1}}
\sum_{i=1}^n r_i^\ell Y_{\ell m}(\theta_i,\phi_i)$
where 
 ($r_i,\theta_i,\phi_i$) are the polar coordinates of the ith ion with respect to the center-of-mass
 of the cluster and $r_s$ is the density parameter of bulk Na.
We defined a parameter 
$\Delta_{\ell}$$=
$$\sum_{\ell'=2}^\ell\sum_{m=-\ell'}^{\ell'} |a_{\ell',m}|^2$ 
that measures the deviation from sphericity due to the multipole
modes up to the order $\ell$. 
Fig.~2e shows a remarkable correlation ($R^2$$=$84\%, see ref.~\cite{R2})
between $\Delta^t_{\ell=5}$ and the width 
$\Gamma^t$ of the 
spectrum at time $t$ 
($\Gamma^t$ is twice the standard deviation of 
$\sigma^t(\omega)$  for  $2 \leq\omega\leq 3.5$ eV).
Furthermore, including in such analysis only 
quadrupole and octupole deformations (i.e. $\Delta^t_{\ell=3}$) 
results in a similar high value of $R^2$(73 \%), 
while limiting consideration to the quadrupole deformation 
alone (i.e. $\Delta^t_{\ell=2}$) yields a poor correlation
($R^2$$=$19 \%) pointing to the strong influence of octupolar 
deformations on the optical response of metal clusters (even closed shell
ones i.e. Na$_9^+$).
This analysis confirms our conclusion pertaining to the 
importance of the isomerization mechanism (III) in explaining the 
thermal evolution of photoabsorption spectra (particularly 
at high temperatures).

%
%  Conclusions
%
In summary, we demonstrated that calculations of absolute photoabsorption 
cross sections using the TDLDA in conjunction with {\it ab initio} MD simulations 
allow first-principles quantitative description and interpretation 
of optical spectra measured for sodium cations at various temperatures.
Spectral line fragmentation resulting from symmetry-breaking ionic motions 
that lift electronic level degeneracies, frequency shifts (spectral sweeps)
of the entire spectrum due to symmetry-conserving breathing vibrations, 
and structural isomerizations are identified as the main thermal line-broadening
mechanisms with  the first two operative for the smaller cluster (Na$_3^+$ 
and Na$_5^+$) and all three for larger ones.

This work is supported by the US DOE, the  Deutsche
Forschungsgemeinschaft (MM), and the Academy of Finland (HH). We thank R.N. Barnett
for discussions. Computations
were done on Cray T3E at NERSC Berkeley and NIC J\"ulich.
%
%
%Literature
%

%
\end{multicols}

\begin{thebibliography}{200}
\bibitem{Conv} The convergence
of $\sigma(\omega)$ with respect to 
the size of the computational cell (48 $a_0$ in all three dimensions),
and the number of orbitals (30-60) has been carefully checked.
\bibitem{OrbContrib} For the smallest clusters (Na$_3^+$ and Na$_5^+$) 
an analysis 
of the largest elements in $F_{I,ij}$ (see eq.~\ref{f_I} and 
ref.~\cite{Casida}) allows assignment of the particular particle-hole 
pairs $(i,j)$ that make a dominant contribution  to a certain optical line $I$.
Such analysis is not fruitful for larger clusters whose level spectra are more 
dense.
\bibitem{Na7}
While in the text we focus our discussion on the larger cluster
Na$_9^+$, we remark that isomerization broadening was found 
by us also for sodium Na$_7^+$, where at $T$$=$300~K we observed 
structural transitions between the D$_{5h}$ pentagonal bipyramidal
 GS and a D$_{6h}$ planar hexagonal isomer (lying 141 meV above 
the GS) resulting in a bimodal spectral maximum for 
$\omega$$\sim$ 2.4 eV, in addition to a broad absorption peak 
at 3.5 eV (see ref.~\cite{Schmidt}).
\bibitem{Na9GS} The Na$_9^+$ GS structure  (a tricapped  trigonal
prism, $\alpha$ in Fig. 1d) is lower by 98 meV than the Na$_9^+$ GS 
reported in ref.~\cite{Bonacic} and depicted as configuration $\beta$ on 
the left of Fig. 1e (a bicapped bipyramide with a C$_{2v}$ 
symmetry).
\bibitem{boltzman} We note that attempts to describe the simulated 
450 K spectrum shown in Fig. 1e by Boltzmann-weighted averaging of
the broadened static 0 K spectra of the $\alpha$ and $\beta$  
isomers (insets in Figs. 1d and 1e) do not lead to
satisfactory agreement with our ab-initio results and the experimental
data, unless unphysically large broadening
parameters are used. Furthermore, this procedure fails to describe
the observed  shift of the high-temperature 
spectrum to lower energies due to thermal
expansion of the cluster.
\bibitem{sweeping}
From our simulations of Na$_9^+$ at 450 K, we observed a direct correlation 
between the ionic repulsion energy and the mean absorption frequency
(similar to that shown in  Fig.~2d for Na$_5^+$). The line broadening due 
to the breathing mode (mechanism II) implied by such correlation,
contributes merely 66 meV to the total width ($\sim$0.6 eV, see Fig.~1e).
\bibitem{R2} The coefficient 
$R$$=$$(1/t_s)\int dt (x(t)-\bar x) (y(t)-\bar y)/(s_x s_y)$ is a common measure of the correlation between two time series $x(t)$ and $y(t)$ with means  $\bar x$ and $\bar y$
 and standard deviations $s_x$ and $s_y$,  respectively.
\end{thebibliography}
\end{document}